# DIGITAL SIGNAL PROCESSSING FUNCTIONS FOR ULTRA- LOW FREQUENCY CALIBRATIONS

Henrik Ingerslev[1], Søren Andresen[2], Jacob Holm Winther[3]

[1] Brüel & Kjær, Danish primary laboratory for acoustics, Denmark, henrik.ingerslev@bksv.com
[2] Hottinger, Brüel and Kjær, Denmark, soren.andresen@bksv.com
[3] Brüel & Kjær, Danish primary laboratory for acoustics, Denmark, jacobholm.winther@bksv.com

**Abstract:**
The demand from industry to produce accurate acceleration measurements down to ever lower frequencies and with ever lower noise is increasing [1][2]. Different vibration transducers are used today for many different purposes within this area, like detection and warning for earthquakes [3], detection of nuclear testing [4], and monitoring of the environment [5]. Accelerometers for such purposes must be calibrated in order to yield trustworthy results and provide traceability to the SI-system accordingly [6]. For these calibrations to be feasible, suitable ultra low-noise accelerometers and/or signal processing functions are needed [7].

Here we present two digital signal processing (DSP) functions designed to measure ultra low-noise acceleration in calibration systems. The DSP functions use dual channel signal analysis on signals from two accelerometers measuring the same signal and use the coherence between the two signals to reduce noise. Simulations show that the two DSP functions are estimating calibration signals better than the standard analysis.

The results presented here are intended to be used in key comparison studies of accelerometer calibration systems [8][9], and may help extend frequency range down to ultra-low frequencies of around 10mHz.

## 1. INTRODUCTION

In the field of dual channel signal analysis there are some very powerful functions for analysing signals, such as the well-known frequency response function and coherence. But there are also other functions like the coherent power function (COP) and non-coherent power function which are very powerful for decomposing noisy signals into the coherent part and the non-coherent part [10][11][12]. Consider an accelerometer calibration setup with two accelerometers mounted close to each other and measuring the same signal. They will both measure the acceleration of the shaker, but since they are different sensors with different conditioning, they will have different noise. The two signals will have a coherent part which is the acceleration signal and a noncoherent part which is the noise. Hence, the COP can be a powerful tool for extracting the signal from the noise and thereby increase the measuring accuracy of the power.

A similar function for increasing the measuring accuracy of the phase by separating the coherent phase from the non-coherent phase is also derived in the next section, called the coherent phase (or argument) function (COA). For the COA to work in a proper manner, it is crucial that the signal applied to the shaker is a continuous signal, like a sine or a multi-sine, and that the frequencies of the sines are very precise and phase synchronized with the frequencies of the Fourier transformation, to prevent the phase from drifting or even make jumps. More details on this will be given in section 2.2.

The two DSP functions analysed in this article may prove relevant to be used for e.g. key comparison of calibration systems down to extremely low frequencies of around 10mHz where noise becomes a real challenge [7][8][9].

The degree to which the COP, and the COA can separate a signal into coherent and noncoherent parts increases with the length of the measurement, and generally depends on parameters like how many time-samples the measurement is divided into, how long each time-sample is, the sampling rate, and the Fourier transform used.

## 2. DIGITAL SIGNAL PROCESSING FUNCTIONS - THEORY

In this section the theory for two DSP functions is outlined. The two functions are based on dual channel signal analysis and can give a better estimate of signals in very noisy environments, than standard signal analysis. The first function is the COP for estimating the power or amplitude of the signal. And the second is the COA for estimating the phase. Both functions rely on the coherence between the two signals.

### 2.1. Coherent power function

Consider two sensors both measuring the same stimuli and positioned close enough for their mutual transfer function to be considered unity. As illustrated in Figure 1 the signal without noise called $u(t)$, and the noise from each sensor called $n(t)$ and $m(t)$ yields the output signals from the two sensors, called $a(t)$ and $b(t)$.

Now consider $j = 1 \dots N$ discrete time-samples measured with the two sensors:

$$a_j(t_i) = u_j(t_i) + n_j(t_i) \qquad (1)$$

$$b_j(t_i) = u_j(t_i) + m_j(t_i) \qquad (2)$$

Here $t_i$ is the discrete time in each time-sample, $u_j$ is the discrete signal, and $n_j$ and $m_j$ are discrete noise in each time-sample. By discrete Fourier transformation of equation (1) and (2) we get

$$A_j(f_k) = U_j(f_k) + N_j(f_k) \qquad (3)$$

$$B_j(f_k) = U_j(f_k) + M_j(f_k) \qquad (4)$$

where $f_k$ is the discrete frequency, $U_j$, $N_j$, and $M_j$ are the discrete Fourier transforms of $u_j$, $n_j$, and $m_j$ respectively.

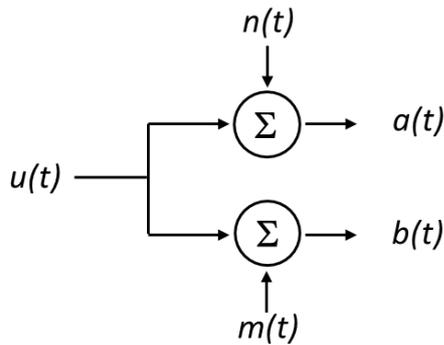

*Figure 1: Illustration of the signals used in the dual channel signal analysis and derivation of the two DSP functions. $u(t)$ is the signal we want to measure, and $n(t)$ and $m(t)$ are the noise contributions from each sensor, which then yields the two output signals $a(t)$ and $b(t)$.*

The cross spectrum is given by

$$S_{AB}(f_k) = \frac{1}{N} \sum_{j=0}^{N-1} A_j(f_k) B_j^*(f_k) \qquad (5)$$

for $N \to \infty$, and * denotes complex conjugate.

By inserting equation (3) and (4) in equation (5), and using that $U_j$, $N_j$, and $M_j$ are uncorrelated the cross spectrum can be given by

$$S_{AB}(f_k) = \frac{1}{N} \sum_{j=0}^{N-1} U_j(f_k) U_j^*(f_k) \stackrel{\text{def}}{=} S_{UU}(f_k) \qquad (6)$$

where $S_{UU}(f_k)$ is the power spectrum of the signal without noise, i.e. the coherent power. Therefore, the COP can in this setup be given by:

$$COP(f_k) = S_{AB}(f_k) \qquad (7)$$

### 2.2. Coherent phase function

Consider the following function, for $N \to \infty$:

$$D_{AB}(f_k) = \frac{1}{N} \sum_{j=0}^{N-1} A_j(f_k) B_j(f_k) \qquad (8)$$

It looks like the cross spectrum from equation (5), but without the complex conjugation. This "non-conjugated cross spectrum" is very useful for deriving a function for measuring the phase of the coherent signal.

We can similarly to the derivation of the COP insert equation (3) and (4) in (8), and use that $U_j$, $N_j$, and $M_j$ are uncorrelated. Hence, the non-conjugated cross spectrum can now be given by, where we have omitted the $f_k$ dependence to make room.

$$D_{AB} = \frac{1}{N} \sum_{j=0}^{N-1} U_j U_j \qquad (9)$$

$$= \frac{1}{N} \sum_{j=0}^{N-1} |U_j|^2 \exp(2i \angle U_j) \qquad (10)$$

$$\simeq \frac{1}{N} \sum_{j=0}^{N-1} |U_j|^2 \exp\left(2i \frac{1}{N} \sum_{j=0}^{N-1} \angle U_j\right) \qquad (11)$$

$$= S_{UU} \exp(2i \overline{\angle U}) \qquad (12)$$

From equation (10) to (11) we have approximated the summation of vectors with length $|U_j|^2$ and angle $2\angle U_j$ by vectors with correct length but all with the mean angle $2\overline{\angle U}$. Figure 2 illustrates this approximation and shows that for relatively small changes in angle from sample to sample this is a good approximation. In calibration applications the signal measures the acceleration of the shaker. And by applying a sine or multi-sine with frequencies exactly the same and phase synchronized with the Fourier frequencies to the shaker, the phase from sample to sample can be kept steady without drifting and the approximation will therefore be very good in such calibration applications.

In equation (12) $\overline{\angle U}$ is the mean phase of the coherent signal. Therefore, the coherent phase function can be given by $COA = \overline{\angle U}$. And by rewriting equation (12), the COA can be given by:

$$COA(f_k) = \frac{1}{2} \text{Imag}\bigl(\ln(D_{AB}(f_k))\bigr) \qquad (13)$$

We have in the derivation of equation (13) used that the signal power $S_{UU}$ is purely real.

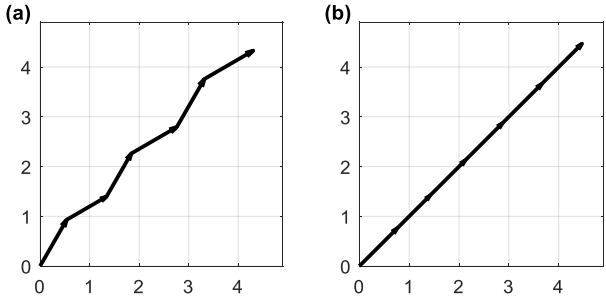

Figure 2: schematic illustration showing the approximation from equation (10) to (11). (a) shows the correct summation where each arrow has length $|U_j|^2$ and angle $2\angle U_j$ as in equation (10), and (b) shows the approximative summation where each arrow has correct length but a mean angle $2\overline{\angle U}$ as in equation (11). As can be seen from (a) to (b), if $U_j$ does not change to much between samples the approximation is good.

## 3. SIMULATIONS

In this section we test the COP and the COA functions on simulated data. We calculate the discrepancy between the functions estimate of the signal amplitude and phase, and the true values. And we compare this with standard signal analysis which is to average the amplitude and phase over all the samples.

When measuring the cross spectrum or the non-conjugated cross spectrum for finding the COP and the COA we only measure a finite number of samples $N$, which therefore only yields an estimate of the COP and the COA. Hence, the more samples the more precise the estimate will be, and in the following the simulations is based on a 102400s (~28h) time sample divided into $N = 1024$ samples of 100s each, with 2048 discrete measurement points in each sample, and a Fourier transform from 10mHz to 10.24Hz with a 10mHz step.

How well the COP and COA works is estimated by representing $a_j$, and $b_j$ from equation (1) and (2) with simulated data. Here the signal $u_j$ is a multi-sine with $l = 1 \ldots M$ frequencies $(f_l)$ and phases $(\phi_l)$, all with amplitude $U0$:

$$u_j(t_i) = U0 \sum_{l=1}^{M} \sin(2\pi f_l t_i + \phi_l) \quad (14)$$

The noise $n_j$, and $m_j$ are random generated white noise with amplitude $N0$. And the signal to noise ratio is given by:

$$SNR = \frac{U0}{N0} \quad (15)$$

The frequencies $(f_l)$ of the multi-sine in equation (14) must be precisely the same or synchronized to a subset of the Fourier transformation frequencies $(f_k)$ from equation (3) and (4), otherwise the COA will drift. This requirement is easily met in the simulations presented here, since the subset of frequencies $(f_l)$ can be set to be identical to some of the frequencies $(f_k)$. But in real measurements this requirement might be challenging to meet.

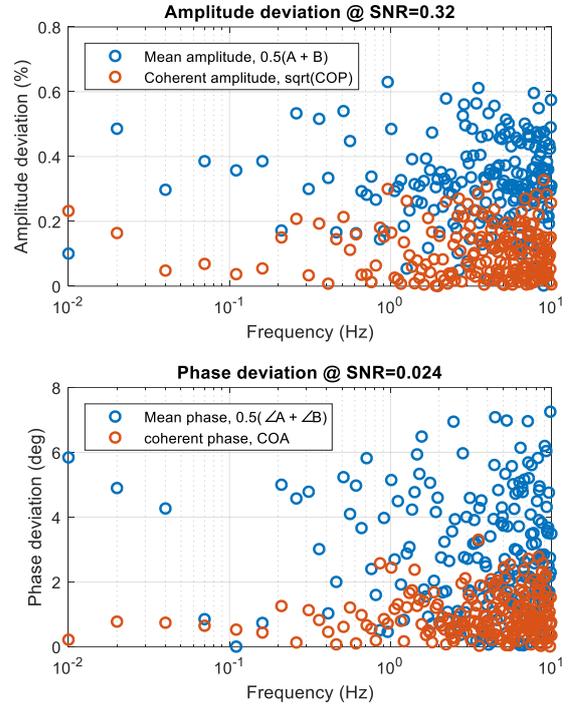

Figure 3: Based on simulated data the coherent power function and coherent phase function is tested for its strength for estimating the amplitude and phase of sine waves in noisy data. The graph shows the deviation of the amplitude and phase from the true vales. (a) shows the mean amplitude, $0.5(A + B)$, and the coherent amplitude, $\sqrt{COP}$. (b) shows the mean phase $0.5(\angle A + \angle B)$, and the coherent phase COA.

Figure 3(a) shows the discrepancy between the signal amplitudes $U0$ from equation (14) and the coherent amplitude, defined as $\sqrt{COP(f_k)}$, for a signal to noise ratio of $SNR = 0.32$ in red circles. And for comparison we also plot the mean amplitude in blue circles, that is, $0.5(A(f_k) + B(f_k))$ where $A(f_k) = \sum|A_j(f_k)|$ and $B(f_k) = \sum|A_j(f_k)|$ is the average amplitudes over all samples. And we plot only at the frequencies of the signal, i.e. at $f_k = f_l$. It is clearly seen that the COP function estimates the amplitude better that the mean amplitude in the full frequency range.

Similarly, Figure 3(b) shows the discrepancy between the phase $\phi_l$ and the COA for a signal to noise ratio of $SNR = 0.024$ in red circles. And for comparison we also plot the mean phase defined as $0.5(\angle A + \angle B)$ where $\angle A = \sum \arg(A_j(f_k))$ and $\angle B = \sum \arg(B_j(f_k))$ are the average phases over all samples. And it is seen that the COA estimates the phase better that the mean phase.

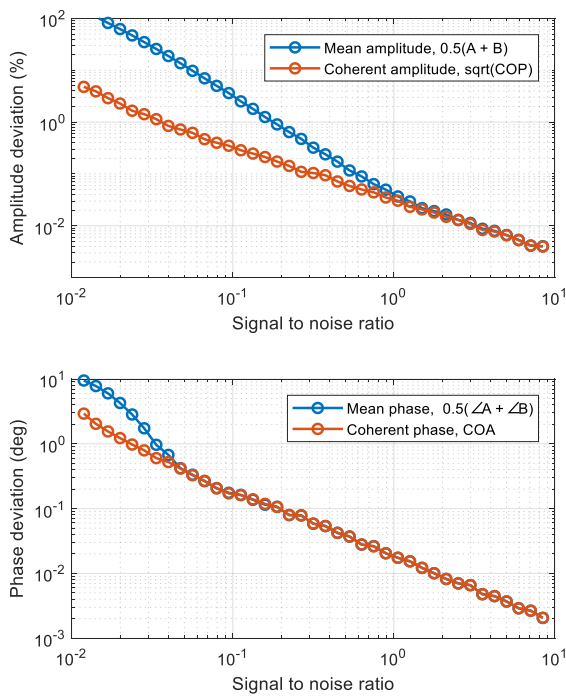

*Figure 4: Simulated deviation in amplitude and phase versus signal to noise ratio, in (a) for the mean amplitude and the coherent amplitude defined as $\sqrt{COP}$, and in (b) for the mean phase and the coherent phase COA. Each deviation point plotted here is an average over all the frequencies from Figure 3.*

The deviations plotted in the four curves in Figure 3 seems to be independent of frequency. Therefore, by averaging the deviation for each curve in Figure 3 over all the frequencies $f_l$, we get a mean deviation for the COP and the mean amplitude at $SNR = 0.32$, and a mean deviation for the COA and the mean phases at $SNR = 0.024$. We have done this for a range of signal to noise ratios from $SNR = 0.01$ to $SNR = 10$ and plotted it in Figure 4. It shows that the COP is better than the mean amplitude from about $SNR = 1$, and the COA is better than the mean phase from about $SNR = 0.04$. The data in Figure 4 depends highly on the length of the time sample and Fourier transform used.

## 4. SUMMARY

We have derived two DSP functions for very accurate measurements of amplitude and phase from two accelerometers measuring the same signal. We have tested the two DSP functions on simulated data and our findings based on the simulations shows promise to the functions as good tools for accurately measuring amplitudes and phases of a multi-sine wave in a noisy environment.

These findings may prove useful for key comparisons of accelerometer calibration systems down to ultra-low frequencies, since for such measurements noise becomes a huge problem as the frequencies approaches 10mHz. Hence, by replacing the accelerometer used in key comparisons by two accelerometers and by using the two DSP functions described here, the frequency range in key comparisons may be possible to extend down to ultra-low frequencies of around 10mHz.

## 5. REFERENCES

[1] T. Bruns, S. Gazioch: "Correction of shaker flatness deviations in very low frequency primary accelerometer calibration, IOP Metrologia, vol. 53, no. 3, pp. 986 (2016).

[2] J. H. Winther, T. R. Licht, "Primary calibration of reference standard back-to-back vibration transducers using modern calibration standard vibration exciters", Joint Conference IMEKO TC3, TC5, TC22, Helsinki, (2017).

[3] G. Marra, C. Clivati et al., "Ultra-stable laser interferometry for earthquake detection with terrestrial and submarine optical cables", Science, vol. 361, Issue 6401, pp. 486-490, (2018).

[4] P. Gaebler, L Ceranna et al., "A multi-technology analysis of the 2017 North Korean nuclear test", Solid Earth, 10, 59–78, (2019).

[5] R. A. Hazelwood, P. C. Macey, S. P. Robinson, L. S. Wang, "Optimal Transmission of Interface Vibration Wavelets - A Simulation of Seabed Seismic Responses", J. Mar. Sci. Eng. 6, 61-79, (2018).

[6] L. Klaus, M. Kobusch, Seismometer Calibration Using a Multi-Component Acceleration Exciter, IOP Conf. Series: Journal of Physics: Conf. Series 1065 (2018).

[7] EU EMPIR project "Metrology for low-frequency sound and vibration", (2020).

[8] BIPM Key comparison, CCAUV.V-K5 (2017), https://www.bipm.org/kcdb/comparison?id=453

[9] EURAMET Key comparison, AUV.V-K5 (2019), https://www.bipm.org/kcdb/comparison?id=1631

[10] H. Herlufsen, "Dual channel FFT analysis part I", B&K Technical Review, no.1, (1984).

[11] J. S. Bendat, A. G. Piersol, "Random Data", Wiley-Interscience, (1986).

[12] J. S. Bendat, A. G. Piersol, "Engineering applications of correlation and spectral analysis", Wiley, New York, (1993).